# Image Authentication using Visual Cryptography

Rahul Saranjame (201301101)

*DA-IICT, Gandhinagar*

`201301101@daiict.ac.in`

*Supervisor*
*Dr. Manik Lal Das*

*Abstract* – **This report gives a novel technique of image encryption and authentication by combining elements of Visual Cryptography and Public Key Cryptography. A prominent attack involving generation of fake shares to cheat honest users has been described and a demonstration of the proposed system employing a centralised server to generate shares and authenticate them on the basis of requests is made as a counter to the described attack.**

## I. INTRODUCTION

### A. Overview

Visual Cryptography was first introduced by Naor and Shamir where they divided a secret image into 'n' shares using a codebook which basically prescribed what each pixel in the share would look like according to the share in the secret image. This system had a major drawback of pixel expansion where each pixel in the secret image was mapped to an nxn pixel leading to expansion of the shares. Further, the requirement of the codebook was another downside that this technique had. The main advantage of Visual Cryptography as an encryption technique is that it renders any computation for decryption unnecessary. The only computational cost involved is in the encryption phase whereas decryption is simply carried out by the Human Visual System by superimposing the shares.

Kafri and Keren first proposed the method of encryption using Random Grids. This eliminates the two drawbacks of the earlier scheme whilst retaining its computational simplicity. The secret image is broken down into noise like grids called shares that are the same size as the image. This eliminates the need for a codebook and does not require pixel expansion. The decryption process occurs by simple superimposition.

Chen and Tsao further proposed the (2,n) and (n,n) secret sharing scheme based on Random Grids. In (2,n) RG based method, the image is broken down into 'n' random grids and only on the acquisition of 2 or more shares can it be reconstructed. In (n,n) RG based method, all 'n' need to be acquired and superimposed to recreate the secret image.

They then came up with the novel method of (k,n) secret sharing which is what this project primarily uses as a basis. The principle remains the same: on getting atleast 'k' out of the 'n' shares the image can be recovered by simple visual detection but having 'k-1' or less shares reveals absolutely nothing about the secret image.

### B. Real World Context

One of the major fields where visual cryptography has applications is in medical sciences where patient data confidentiality is of prime importance. Assuming a typical hospital scenario where patient records are stored on a central database and there is need for strict confidentiality, this project makes use of the (k,n) secret sharing technique to achieve the purpose. Each patient may have a number of scans and image records that are supposed to be available only to a certain number of doctors along with the patient. The technique described here incorporates elements of Public Key Cryptography with Visual Cryptography to thwart the attack later described in the report. The main reason that the attack is successful in exploiting the vulnerability existing in the scheme is due to a lack of authentication. Once such a system is introduced the attack can be successfully circumvented.

Further a lot of computational overhead can be reduced by adopting the scheme discussed in the report and it also has the added benefit of preventing attacks that Public Key Cryptography is secure against. Using elements from both domains ensures that maximum security is provided, integrity is maintained and computational cost for decryption is negligible.

The report is organised as follows. Section 2 gives a brief insight into the way shares are developed in (2,n) and (k,n) RG based Share Generation and describes why in the considered context of healthcare the need for share generation arises and how it is carried out. Section 3 elucidates in detail the attack that exploits vulnerabilities in the considered (2,n) and (k,n) RG based VC. It also discusses the practical problems of such an attack being successful and the damage caused. Section 4 provides a detailed illustration of the proposed system for preventing the attack and expounds upon the practical applications of the new scheme. Section 5 concludes the report.

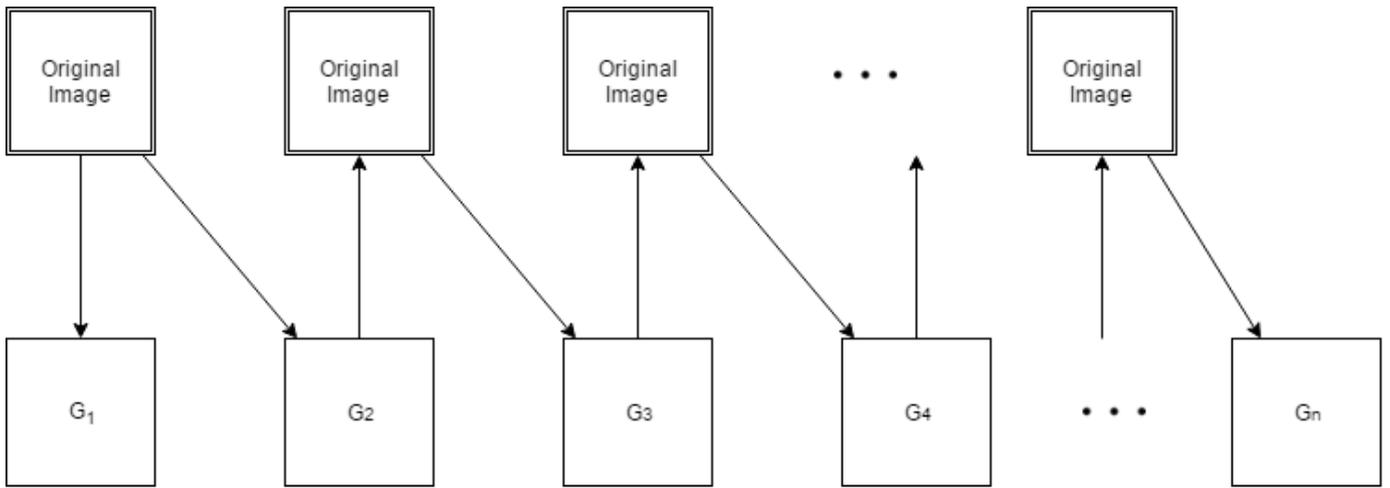

Fig. 1

## II. SHARE GENERATION

This section elaborates on the process of share generation in (2,n) and (k,n) Random Grid VC.

### A. (2,n) Random Grid Share Generation

To generate 'n' shares, such that any 2 are required to recreate the original image, the first share $G_1$ is randomly generated. The next share $G_2$ is generated such that superimposing it with $G_1$ gives S. Now, $G_3$ is also generated such that superimposing with $G_1$ or $G_2$ yields the original image S. The process is repeated till all 'n' shares are generated.

Irrespective of the original image S we generate $G_1$ by assigning random values 0 or 1 at each pixel location. For each corresponding pixel in the subsequent shares the following algorithm is used.

**Algorithm 1:**
**Input:** Original Image S, Randomly generated $G_1$
**Output:** $G_x$ (x=2,3…..n)

Function CreateShare(S, $G_1$)
for i: 1 to r
  for j: 1 to c

  if, S(i,j) = 0 (white pixel in original image)
  Gx(i,j) = G1(i,j)

  else if, S(i,j) = 1 (black pixel in original image)
  Gx(i,j) = rand(0,1).

  end
end

Rand(0,1) is the function that randomly assigns a value 0 or 1.

To ensure genuine randomness of the rand() function we can take a generate a large number of the order of $10^7$ and then take remainder on being divided by 2 as a crude measure.

Figure 1 diagrammatically represents (2,n) RG based Share Generation.

### B. (k,n) Random Grid Share Generation

The process of generating shares in (k,n) is different. The original image is used to generate $G_1$ and $IG_1$ which is an intermediate grid. Now $IG_1$ is used as an image and $G_2$ and $IG_2$ are generated. This process continues till $IG_{n-2}$ is split into $G_{n-1}$ and $G_n$. $G_1$ is generated randomly and $IG_1$ is so generated that superimposition with $G_1$ gives the original image. Thus, subsequent shares are generated by treating the intermediate grids as images (as shown in Figure 2).

### C. Applications of Random Grid VC

Extending the use of Random Grids to our example, it is easy to generate shares for patient records. The required scans and metrics for each patient are obtained and converted into a 2 dimensional matrix. The X-rays and other scans initially undergo image processing to convert them into an appropriate grayscale format for depicting as 2 dimensional matrices. These are then used as the original image and divided into multiple shares which are distributed among the concerned doctors and one share is retained by the patient.

## III. ATTACK DESCRIPTION

This section briefly describes the attack that can be conducted in Random Grid based Visual Cryptography.

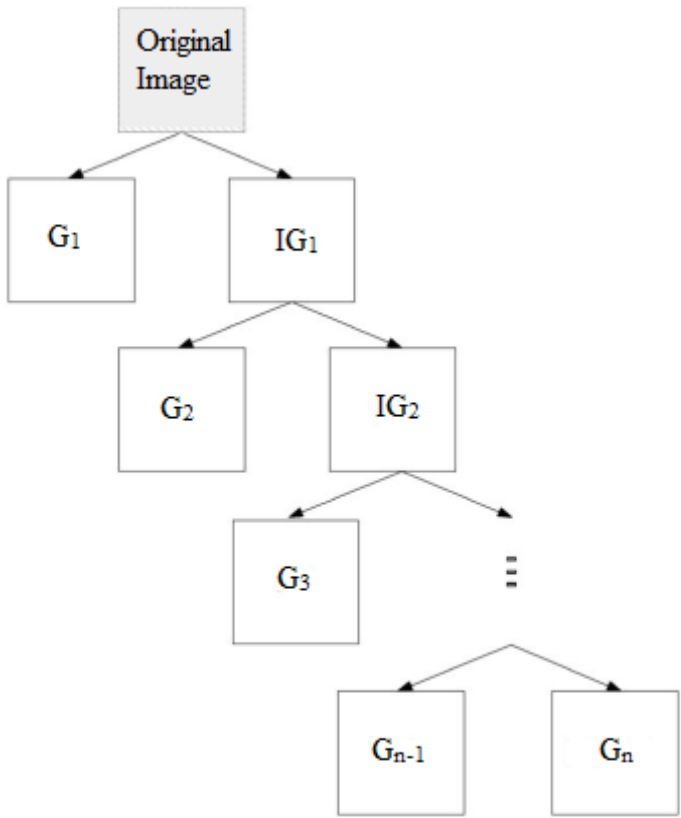

Fig. 2

### A. Attack Mechanism in (2,3) Random Grid VC

The initial step involves breaking down the secret image into 3 random grids by using the algorithm given by Chan and Tsao for (2,n) Random Grid based Visual Secret Sharing. Let us assume that these random grids or shares are given to X, Y and Z. Further, let X and Y be the malicious users who wish to cheat honest Z. According to the principle of (2,3) Visual Cryptography, X and Y can superimpose their shares and reconstruct the secret image. The next step involves creating a fake image by altering the original secret image. Let us call this fake image FI. FI is then used to create a Fake Grid, FG that is circulated as a share. When the honest user Z uses FG as a share to recreate the image he does not get the original image but the fake image.

The fake image is obtained by altering some of the white pixels of the original image to black.

Supposing X's share is $G_1$, the following algorithm is used to compute each pixel of the Fake Grid FG.

Figure 3 depicts the generation of the Fake Grid.

**Algorithm 2:**
**Input:** Reconstructed Image I, Fake Image FI, X's Share $G_1$

**Output:** Fake Grid FG

Function FakeGrid(I, FI, $G_1$)
for i: 1 to r
  for j: 1 to c

  if, FI(i,j) = I(i,j)
  FG(i,j) = $G_1$(i,j)

  else,
  FG(i,j) = rand(0,1).

  end
end

### B. Attack Mechanism for general (k,n) Random Grid VC

The basic attack methodology remains quite similar to the (2,3) random grid VC described above. The difference lies in the fact that now the number of shares required to recreate the image is 'k' which means that given 'k' malicious users they can superimpose their shares and obtain the original image. They then proceed as earlier to generate fake grids that give an incorrect image on superimposition. The major difference is the requirement of 'k-1' fake shares that are required for the legitimate user to be cheated.

### C. Threat in the real world

Furthering the aforementioned scenario of the application in the field of medical sciences let us analyse the attack that the system is susceptible to. Suppose the patient has important image data that cannot be exposed and is only privy to a select number of doctors the pre-discussed attack can be used for a variety of nefarious uses. Let us assume each of the 'n-1' (one share is retained by the patient) doctors are given a share and that on obtaining 'k' shares they can recreate the brain scan of the patient and analyse if for possible tumours. Supposing 'k' dishonest doctors come together and decide to fabricate a fake image that shows the presence of a tumour in an otherwise healthy brain they can easily do so. They simply recreate the brain scan, alter it and create fake shares. These on superimposition give the brain scan that shows false presence of a tumour. Such tampering of data is highly deleterious for the patient who will be forced to take medication that is not required. Further even when any of the other honest doctors use these fake shares they will obtain a tumorous brain scan and give faulty medical opinion to the patients. The major strength of this attack lies in the inability to trace the origin to its perpetrator. Once the fake shares are generated no one can see whether they have been tampered with or not and it is impossible to check who is actually an innocent user and who has malicious intent (Figure 2 shows how a fake image can be recreated by using Fake Grids).

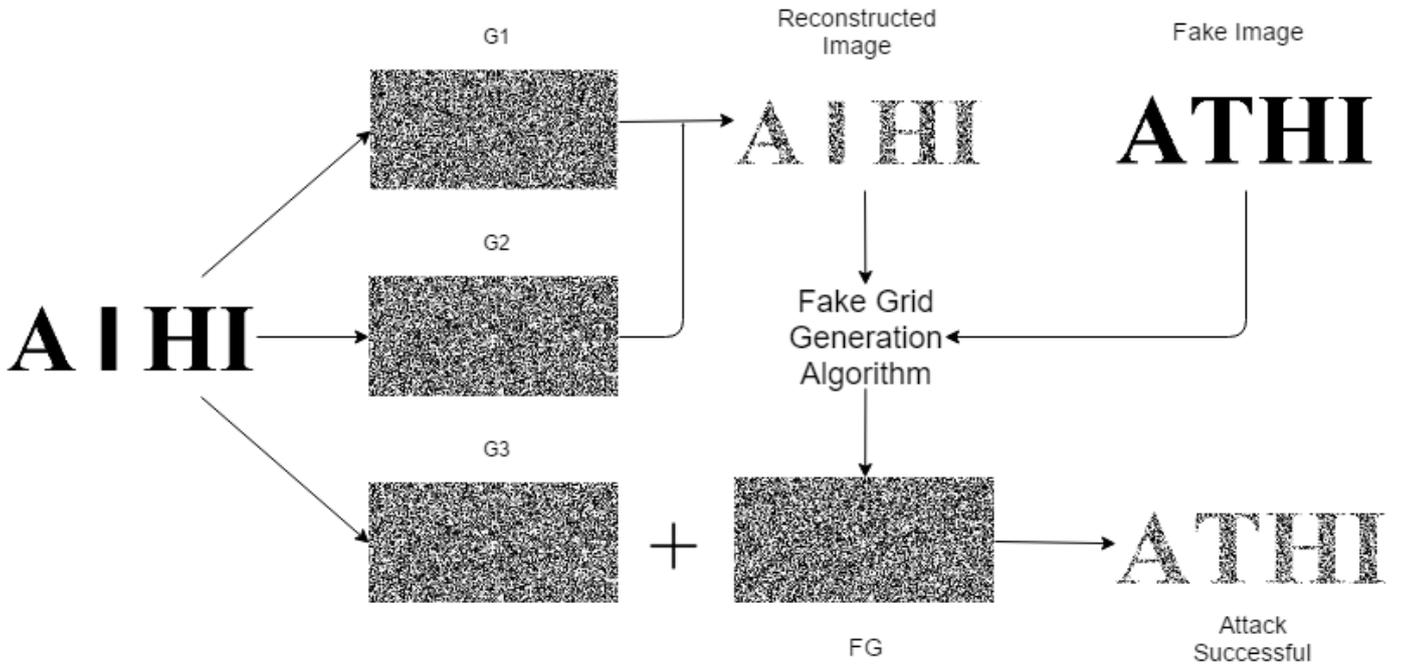

Fig. 3

## IV. THE PROPOSED PREVENTION MECHANISM

A new protocol that combines elements of Public Key Cryptography with Visual Cryptography is proposed. In an elementary way the technique derives its functionality from SSL-TLS protocols.

### A. SSL-TLS: An Overview

SSL-TLS are cryptographic protocols that are widely used in a host of applications such as email and web browsing due to the provision of a secure communication channel over a network. These have three main properties:
1. The data transmitted between the concerned parties is encrypted using Symmetric Key Encryption. The encryption algorithm and the keys are decided before the actual transmission of data starts in something called the handshake phase.
2. The integrity of each message thus communicated can be vouched for due to the use of a MAC (Message Authentication Code) that can be used to detect the loss or modification of data.
3. The identity of the parties communicating is authenticated using Public Key Cryptography.

### B. Share Generation and Distribution

Share Generation is as described in Section 2 of the report. To make the data available to 'n' users but disallow recreating the data to fewer than 'k' users the image is split into 'n' shares and distributed. There exists a central server that has the sole function of generating shares and using a cryptographic hash function to compute digests for each of these shares. The server retains a copy of all these digests for each of the shares and additionally forwards a share and its respective digest to each user.

### C. Authentication and Image reconstruction

Let us assume that user A wants to send his share to B so that B can combine their shares and obtain the image (we are considering a (2,n) RG based protocol here). A can certainly not send his share and MAC to B primarily due to the absence of a secure communication channel between them and also because of the fact that there is no way for B to check the integrity of the data. So all communications are routed through the server.

A and S (the server) initially implement a rudimentary handshake protocol like the one followed in SSL-TLS. All handshakes make use of Public Key Cryptography to authenticate the communicating parties and then decide a session key because Symmetric Key Encryption is much cheaper than Public Key Encryption. Once the session key is established all messages are encrypted using that until it expires.

Let the session key established between A and S be $K_{AS}$. A now sends S a message indicating that he wishes to communicate with B.

**A to S: $\{A, B, N_A\}_{K_{AS}}$**

S establishes a session key for A and B, $K_{AB}$, and sends it to A.

**S to A: $\{N_A+1, K_{AB}\}_{K_{AS}}$**

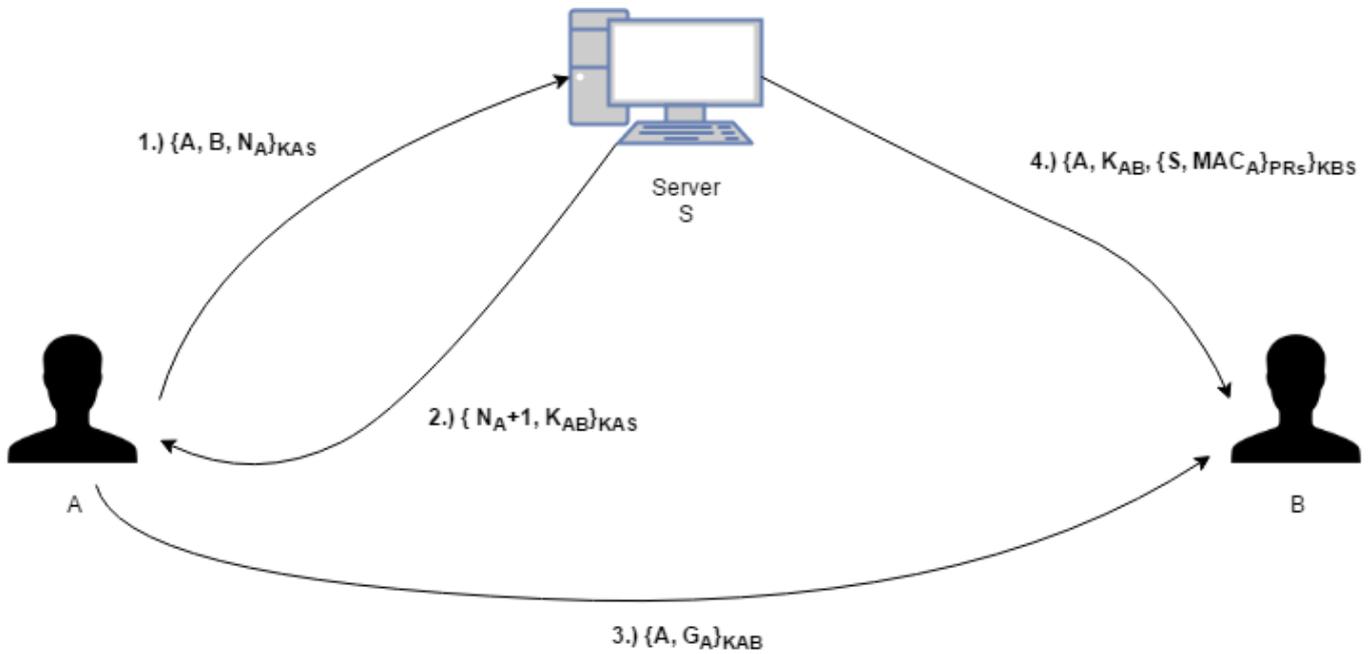

Fig. 4

A now sends his share and MAC value to B for superimposition by encrypting it with $K_{AB}$.

**A to B: $\{A, G_A\}_{K_{AB}}$**

S now runs the handshake protocol with B and establishes a session key with B, $K_{BS}$. Further S provides B with the A-B session key, $K_{AB}$. S also provides B with A's MAC value signed by S's private key for ensuring integrity.

**S to B: $\{A, K_{AB}, \{S, MAC_A\}_{PR_S}\}_{K_{BS}}$**

B now computes the MAC value for the received A's share. Let us call it $MAC^*_A$. B then tallies this computed value with the one signed and sent by the server. If those two are the same it means that A's share has not been tampered with and can be used for superimposition without any threat. B finally retrieves the image using both shares.

Figure 4 pictorially represents the above described protocol.

*D. Security against attacks*

Contextually speaking if 2 ill-intending doctors wish to cheat others in a (2,n) system the attack would be thwarted by the defences in place. We shall see how.
First they will go through the entire process described above and B will manage to recreate the image using A's share (here A and B are both malicious and wish to cheat a third user C). Now according to the attack described they will make use of a fake image in place of the actual image and alter their shares accordingly. Let the altered shares be $G^*_A$ and $G^*_B$.

Now let one of them, say A, want to cheat C. A will first establish a session key with the server and send a request expressing his wish to communicate with C.

**A to S: $\{A, C, N_A\}_{K_{AS}}$**

S will generate a session key for A and C and reply to A.

**S to A: $\{N_A+1, K_{AC}\}_{K_{AS}}$**

Now, A will send the altered share to C encrypted with the session key, $K_{AC}$.

**A to C: $\{A, G^*_A\}_{K_{AC}}$**

Now the server will establish a session key with C and send the MAC value of A's share (this share is unaltered since the server has all original copies) along with $K_{AC}$.

**S to C: $\{A, K_{AC}, \{S, MAC_A\}_{PR_S}\}_{K_{CS}}$**

C will now decrypt the message it received from A and compute the MAC value for $G^*_A$. Let that be represented by $MAC^*_A$.

C now decrypts the message sent by S using their session key $K_{CS}$. He obtains the MAC value for A's share signed by S. C then verifies the signature to ensure S's authenticity and the integrity of the MAC value.

Now, the final stage involves comparing $MAC^*_A$ and $MAC_A$. It is clearly seen that they do not match since A's share was modified. C now knows that the share has been tampered with and does not proceed to recreate the image.

Another added advantage of using this protocol is that because Public Key cryptography is used to establish symmetric keys the cost is lesser. This technique also successfully prevents any MITM (Man in the Middle) attacks.

## V. CONCLUSION

In this report, an innovative technique has been designed to prevent attacks due to generation of fake shares. This technique combines elements of Visual Cryptography and Public Key Cryptography to establish a secure channel for communication between the users thus preventing outside attacks. Further, it also introduces a system of authentication that helps check the integrity of the data sent. No malicious attacker can now exploit the system from the inside or outside.

## ACKNOWLEDGMENTS

I would like to express my heartfelt gratitude to Dr. Manik Lal Das for his constant support and guidance without which this project wouldn't have reached completion.